# Tomonaga-Luttinger liquids and Coulomb blockade in multiwall carbon nanotubes under pressure

M. Monteverde[a],

Centre de Recherches sur les Très Basses Températures, C.N.R.S., BP 166 cedex 09,

38042 Grenoble, France and

Laboratorio de Bajas Temperaturas, Departamento de física, Universidad de

Buenos Aires, Ciudad Universitaria, PAB I, 1428 Buenos Aires, Argentina,

M. Núñez-Regueiro

Centre de Recherches sur les Très Basses Températures, C.N.R.S., BP 166 cedex 09,

38042 Grenoble, France

G. Garbarino

Centre de Recherches sur les Très Basses Températures, C.N.R.S., BP 166 cedex 09,

38042 Grenoble, France and

Laboratorio de Bajas Temperaturas, Departamento de física, Universidad de

Buenos Aires, Ciudad Universitaria, PAB I, 1428 Buenos Aires, Argentina

C. Acha

---

[a] Present address *Service de Physique de l'Etat Condensé (CNRS URA 2464),*

*DSM/DRECAM/SPEC, CEASaclay, 91191 Gif sur Yvette Cedex, France*

Laboratorio de Bajas Temperaturas, Departamento de física, Universidad de Buenos Aires, Ciudad Universitaria, PAB I, 1428 Buenos Aires, Argentina

X. Jing , L. Lu, Z.W. Pan, S. S. Xie

Key Laboratory of Extreme Conditions Physics, Institute of Physics & Centre for the Condensed Matter Physics, Chinese Academy of Sciences, Beijing 10080, People's Republic of China

J. Souletie

Centre de Recherches sur les Très Basses Températures, C.N.R.S., BP 166 cedex 09, 38042 Grenoble, France

and

R. Egger

Institut für Theoretische Physik, Heinrich-Heine-Universität, D-40225 Düsseldorf, Germany

We report that the conductance of macroscopic multiwall nanotube (MWNT) bundles under pressure shows power laws in temperature and voltage, as corresponding to a network of bulk-bulk connected Tomonaga-Luttinger Liquids (LL). Contrary to individual MWNT, where the observed power laws are attributed to Coulomb blockade, the measured ratio for the end and bulk obtained exponents, ~2.4, can only be accounted for by LL theory. At temperatures characteristic of interband separation, it increases due to thermal population of the conducting sheets unoccupied bands.

61.46.+w, 73.63.Fg, 62.50.+p, 75.30.Kz

Individual single wall carbon nanotubes (SWNT) and bundles or ropes containing a few SWNT's are widely accepted to be clear examples of Tomonaga-Luttinger liquids[1,2,3,4] (LL), with tunneling from metallic contacts into the sample following a characteristic temperature dependence as a $T^\alpha$ power law due to a correlation-induced suppression of the density of states near the Fermi level. The exponent $\alpha$ depends on the number $N$ of conducting channels (due to occupied transverse modes in the shells participating in transport), and a parameter $g$ measuring the Coulomb interaction strength. Moreover, it depends on whether one tunnels into the bulk or close to the end of the MWNT, with theoretical predictions[5] given by $\alpha_{bulk} = (g^{-1} + g - 2)/8N$ and $\alpha_{end} = (g^{-1} - 1)/4N$. The multichannel LL character of multiwall nanotubes (MWNT) formed from concentric SWNT of generally incommensurate wrapping is, in contrast, still controversial, since the ballistic nature required for LL behavior is uncertain. The first measurements on untreated individual MWNT's yielded neat ballistic single mode properties[6,7]. However, the majority of other measurements done on individual MWNT's have shown non-ballistic properties[8,9,10,11]. As some carrier back-scattering is expected[12,13], due, e.g., to sample processing or the interactions among the concentric but incommensurate shells, presently a non-LL state characterized by diffusive Altshuler-Aharonov anomalies seems more likely for the low-energy properties of long individual MWNT's. Including environmental fluctuations, such a state can be described by conventional Coulomb blockade (CB) theory. However, for elevated temperatures, even a disordered MWNT is expected to display LL behavior. Measurement of SWNT networks under high pressure [14] have been shown useful to vary the exponent $\alpha$ in a controlled manner, allowing the verification of the theoretically expected correlation between the coefficient of the conductance power law and $\alpha^{-1}$. Here we present high pressure electrical transport measurements performed on several samples of untreated MWNT bundles showing power laws in temperature and voltage. We analyze our data using the main theories that can be responsible for this behavior, i.e. CB and LL.

The MWNT's were synthesized [15,16] through chemical vapour deposition (CVD). High magnification scanning electron microscopy (SEM) analysis shows that the tubules grow out perpendicularly from the substrate and are evenly spaced at an averaged intertubule distance of ~100 nm, forming a highly aligned array. A high-resolution transmission electron microscopy (HRTEM) study shows that most of the tubules are within a diameter range of 20–40 nm. The mean external diameter is ~30 nm with individual MWNT lengths of below one up to $100 \mu m$. A tubule may contain 10-30 walls, depending on its external diameter. The five samples studied here were rather thick (0.07x0.007x0.003cc) untreated bundles of parallel MWNT contacted by four platinum leads (see insert Fig. 2). Considering the cross-sectional area of our experimental setups and an adequate filling factor, there are of the order of 10000 MWNTs per sample. The electrical resistance measurements were performed in a sintered diamond Bridgman anvil apparatus using a pyrophillite gasket and two steatite disks as the pressure medium[17]. The Cu-Be device that locked the anvils could be cycled between 4.2K and 300K in a sealed dewar.

In Fig. 1 we show the temperature dependence of the electrical conductance as a function of pressure for sample A, typical of all measured. We observe three different regimes. At low temperatures and pressures, region **I,** Fig. 1(b), we observe power laws with temperature dependent exponents in the applied bias of the non-linear conductance that have been attributed to CB behavior[14] in samples from the same origin as ours. Above approximately 100K for all pressures, region **III**, Fig. 1(c), we observe an activated regime. In the rest of the measured range, region **II**, we observe a power law $T^\alpha$ dependence. In order to determine the origin of the power laws of region **II**, we have performed a detailed study of the non-linear conductance both in the four wires, 4W(four terminal setup) and crossed configurations, X, see insert Fig. 2. The former probes the whole sample between the voltage leads, while the latter allows the study of solely the platinum lead to nanotube junction. Note that the X configuration has previously been implemented for thin (diameter $< 1\mu m$) MWNT bundles, where CB behavior was found for bad junctions at low temperatures, $0.1K < T < 20K$, and ambient pressure [14]. Thus, focusing on the crossed junction data, we now compare the LL and CB expressions. From LL theory, one finds for the non-linear conductance $G \equiv dI/dV$ [18,19,20]

$$G(V,T) = A\left(\frac{2\pi k_B}{\hbar\omega}\right)^\alpha T^\alpha \cosh\left(\gamma\frac{eV}{2k_BT}\right)\frac{1}{|\Gamma(1+\alpha)|}\left|\Gamma\left(\frac{1+\alpha}{2}+\gamma\frac{ieV}{2\pi k_BT}\right)\right|^2 \quad (1)$$

leading to

$$\frac{G(V,T)}{G(V=0,T)} = \cosh\left(\gamma\frac{eV}{2k_BT}\right)\frac{1}{\left|\Gamma\left(\frac{1+\alpha}{2}\right)\right|^2}\left|\Gamma\left(\frac{1+\alpha}{2}+\gamma\frac{ieV}{2\pi k_BT}\right)\right|^2 \quad (2)$$

while for CB theory[21] the corresponding expression is

$$\frac{G(V,T)}{G(0,T)} = \left|\frac{\Gamma\left(\frac{2+\alpha}{2}+\gamma\frac{ieV}{2\pi k_BT}\right)}{\Gamma\left(\frac{2+\alpha}{2}\right)\left(1+\gamma\frac{ieV}{2\pi k_BT}\right)}\right|^2, \quad (3)$$

where $V$ is the applied voltage, $\Gamma(x)$ is the Gamma function, $k_B$ the Boltzmann constant, $\hbar\omega$ a bandwidth cutoff, $e$ the electron charge, $A$ a constant that includes geometric factors and the parameter $\gamma$ corresponds to the inverse number of the measured junctions weighted by their resistances (see, e.g., Ref.1). We show in Fig. 2 the corresponding fits for sample E at 10.4GPa. Both expressions (2) and (3) give an undistinguishable fit for the data. However, to obtain this fit we need different values of the $\gamma$ parameter for each case. For $M$ junctions in series, depending on the values of the individual junctions, the $\gamma$ parameter must stay within $1/M$ and 1. For the crossed junction, composed of a certain number of single tunnel junctions in parallel [14], we obtain $\gamma = 1 \pm 0.1$ for the LL, but $\gamma = 2.8 \pm 0.4$ for the CB fits in region **II**. As for junctions in series $\gamma \leq 1$ must hold (parallel junctions are not probed due to our normalization convention), only the LL fit gives consistent results, clearly favoring the LL interpretation of the power laws in region **II**. Thus, at high pressure the crossed junction detects a LL behavior and not the CB behavior that has been demonstrated at ambient pressure and at $T < 20K$ in samples from the same origin[14].

To further test the validity of the LL interpretation of region **II**, we analyze the dependence of $G_0(P)$ (defined as $G_0(P) \equiv G(T,P)/T^{\alpha(P)}$) as a function of $\alpha(P)$. This dependence can be extracted[13,15,16,17] from eq. (1) in the limit $eV \ll k_B T$, and takes the form

$$\frac{G_0(\alpha[P])}{A} = \left(\frac{2\pi k_B}{\hbar \omega}\right)^\alpha \frac{1}{|\Gamma(1+\alpha)|}\left|\Gamma\left(\frac{1+\alpha}{2}\right)\right|^2 \qquad (4)$$

By fitting this expression to our data we find that $A$ is sample dependent but that $\omega$ is the same for all the samples. On the lower panel of Fig. 3 we see how all the samples fall on the same curve, which we have chosen to plot as a function of $\alpha^{-1}$, with $\hbar \omega = 6.5 \pm 0.7 eV$ representing the high energy cut-off (bandwidth) of the LL theory.

Though we expect this energy to change with pressure, the experimental error includes this variation and the obtained value roughly agrees with the expectations.

In our 4W configuration at high pressures, we do not expect to be measuring a single MWNT due to the large size of our sample, but a network of interconnected MWNT junctions [13]. Thus, environmental quantum fluctuations of CB are irrelevant for this measurement. For the five different samples, from the LL fit we now extract $\gamma$ values between 0.003 and 0.03, indicating a minimum of ~30 to 300 bulk-bulk MWNT junctions, with an average exponent $\alpha = \alpha_{bulk-bulk} = 2\alpha_{bulk}$, in analogy with what has been measured in SWNT ropes[13]. We argue that our macroscopic sample picks up percolation paths of quasi-ballistic (mean free path $< 1\mu m$) MWNT. At very low temperatures in the Kelvin range, LL power laws should start to saturate due to the finite MWNT length. As those temperatures are within region I, we do not probe this effect. These MWNTs effectively form Tomonaga-Luttinger liquid segments connected by bulk-bulk junctions, as follows from SEM images (Ref. 15) for such samples. It has been shown for SWNT networks that the value of $\alpha_{end}$ can be independently extracted from the power laws corresponding to the particular geometry of the crossed junction[22]. Our device geometry allowed for measurements of $\alpha_{end}$ only in two samples. We find that the ratio $\alpha_{end}/\alpha_{bulk}$ has a value of $2.4 \pm 0.1$, that varies little with pressure within our experimental error, as was also found for SWNT networks[21]. As for all other known mechanisms yielding power laws, this ratio[11,12] should be strictly $= 2$, this measurement again suggests a LL interpretation of region **II**. Furthermore, this ratio allows us to extract the value of the correlation parameter $g = 0.16 \pm 0.02$ from the above expressions for $\alpha_{bulk}$ and $\alpha_{end}$.

On the upper panel of Fig. 3, we plot the variation with pressure of $\alpha_{bulk-bulk}^{-1}$ for five different samples. As for SWNT ropes[13], we observe a linear pressure dependence, though in this case all curves coincide indicating that the dispersion in diameters of our MWNT is smaller than for the SWNT ropes. We can suppose that the verified charge transfer from impurities such as oxygen [23,24] increases with pressure due to the small inter-band gap, increasing the number of conducting channels. If, as usual, in first approximation we assume a charge transfer rate constant with pressure, we obtain from the known band structure for the measured external diameter an excellent fit( a detailed example of this type of analysis in SWNT networks is found in ref. 24) where the only parameter is $dn/dP = 2 \times 10^{-5}\, holes/C/GPa$. This agreement renders alternative scenarios (such as pressure-induced changes in intertube couplings) less likely.

We have stated in the description of Fig. 1 that there are three regions in our $[G,P,T]$ diagram. We have verified that in region **II** we observe LL properties, while region **I** has been already studied in earlier works[14]. In region **I**, due to bad inter MWNT contacts, these devices probably probe a continuous MWNT from lead to lead. In accordance with expectations that a long MWNT should exhibit the effects of disorder[8-12], the device does not yield LL properties, but CB dependencies. This follows from the analysis of Ref.14, employing very similar fitting procedures as done here in region II. As pressure (and temperature) improves the inter-MWNT contacts, carriers choose ballistic nanotube segments, tunneling from one tube to another. This is only possible because we measure a macroscopic sample with a statistically large number of possible percolation paths.

Region **III** shows for all pressures a rounded increase. The shape of the curves is in fact of an Arrhenius type. Due to the small inter-band gaps in MWNT ($E_0 \approx 30 meV$), we can expect a thermally activated channel occupation to be perceptible in MWNT's at temperatures around room temperature. Thus, for large diameter nanotubes, we can expect an increase in the number of conducting bands due to thermal excitation. As $\alpha$ depends inversely on this number, we can predict a variation in $\alpha$ as temperature increases beyond some critical threshold (of the order of the band separation energy $E_0$). So, beyond this temperature at each pressure, we should observe a deviation from a clean power law to a $G(T) = G_0[\alpha(T)] \cdot T^{\alpha(T)}$ regime where the expression for $G_0(\alpha)$ comes from Eq. (4) and

$$\alpha(T,P) = \alpha(T=0,P)\left(1 + \sum_{m=1} 2\exp\left(-\frac{mE_0}{k_B T}\right)\right)^{-1} \qquad (5)$$

where $E_0$ is calculated from the measured nanotube diameter ($\sim 30 nm$), the $\alpha(T=0,P)$ values are obtained from the power law region **II** and by using the two parameters extracted from the previous scaling of expression (3) shown on the lower panel of Fig. 3: $A$ (a constant that includes geometric factors, and varies only from sample to sample) and $\hbar\omega$, that is the same for all samples. In Fig. 1B the solid lines show the calculated $G(P,T)$ expressions, that, in spite of having no adjustable parameter, are in excellent agreement with the measured values. Finally, this interpretation allows us to calculate the surface (not a fit) shown on Fig. 1(a) from the expression for the temperature and pressure dependence of the LL junctions conductance using Eq. (1), Eq. (4) and Eq. (5) with $\hbar\omega = 6.5 \pm 0.7 eV$ and $A = 6 \pm 0.5 E - 3$, that follows the data except in region **I**.

We conclude that devices made from MWNT bundles can change their behaviour from CB[9] for individual MWNT to that corresponding to a network of LL junctions by application of high pressure, i.e. pressure changes the sample from a lead—sample—lead junction to a lead—multi-junction-sample—lead device. We attribute this change to the interconnection of the different MWNT in the bundle under pressure, with percolation path choosing ballistic segments. We verify in region **II** several tests for LL behaviour : power law dependence for the linear conductance $G(T) \propto T^\alpha$, for the non-linear conductance $dI/dV \propto V^\alpha$, the predicted $G(\alpha)$ dependence, and $\alpha_{end}/\alpha_{bulk} = 2.4 \neq 2$. Furthermore, the deviation of the conductance from a power law at high temperatures in region **III** can be naturally accounted for by the thermal population of unoccupied bands

We are grateful C. Berger, H. Bouchiat and F. Lévy for a critical reading of the manuscript. M.M. and G.G. are CONICET from Argentina doctoral fellows. J. Souletie is deceased. R.E. is supported by the DFG-SFB TR 12.

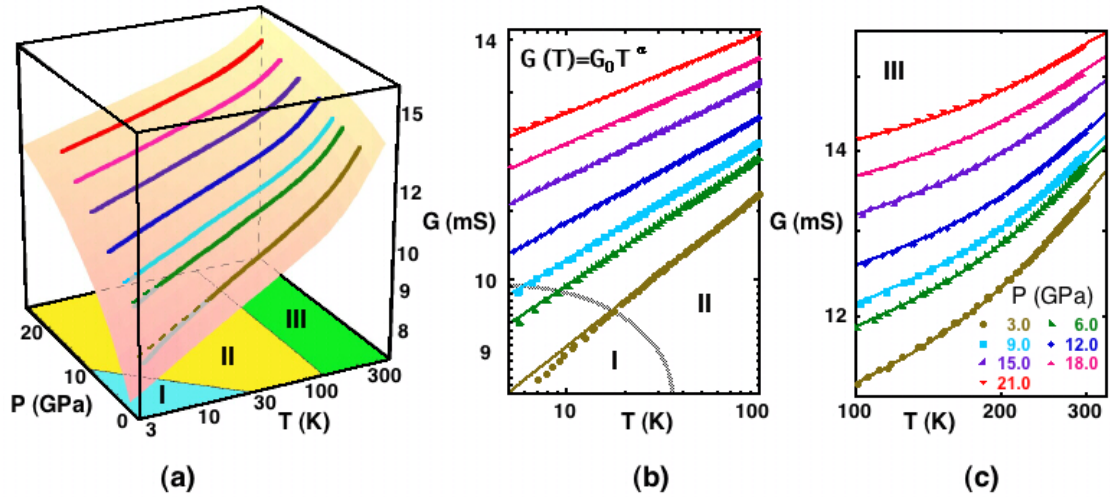

**Figure 1. (Color online)** Temperature dependence of the measured conductance of MWNT sample A at different pressures. (a) 3D representation of the data showing the $(P,T)$ regions corresponding to different behaviors: region I is studied in more detail in Ref. 14, region II show power laws, and region III activated regime. (b) Zoom on the low temperature region to show details of the power law dependence (region II). The low temperature-low pressure bubble indicates the region where the $T^\alpha$ behavior is no longer observed (region I). (c) High temperature zoom showing the activated regime of region III. The curves are described in the text.

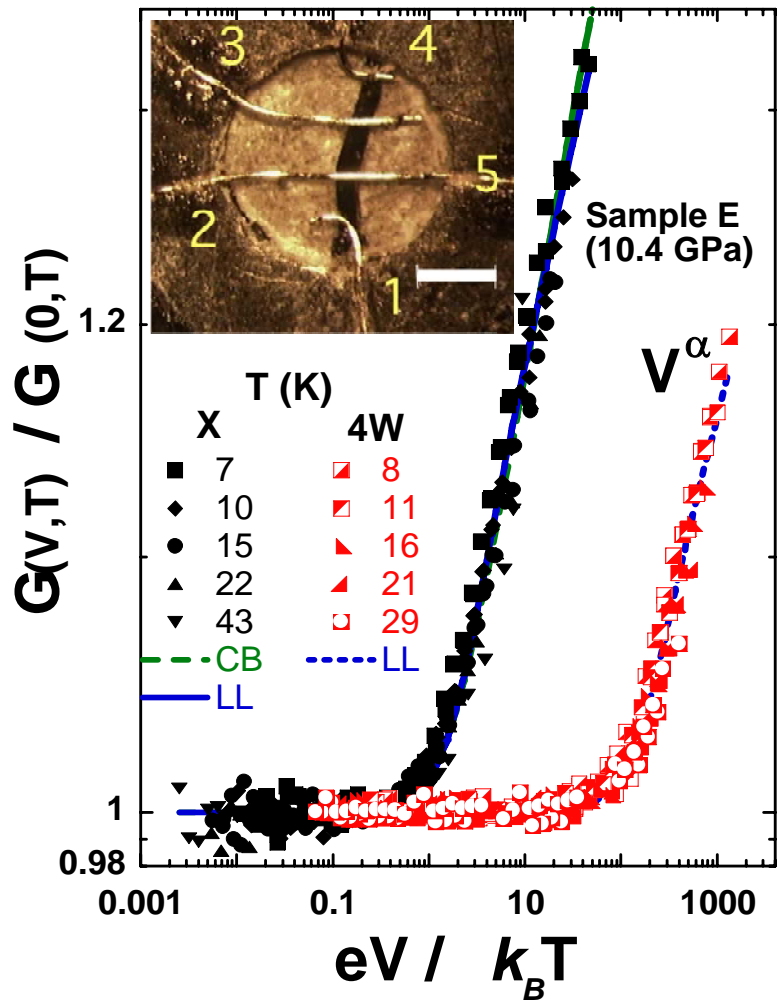

Figure 2 (Color online) Normalized non-linear conductance $dI/dV$ as a function of bias for sample E at 10.4GPa and different temperatures as indicated in the figure. Insert: photograph of a typical sample mounting , the white bar corresponds to $200\mu m$. The crossed (×) configuration uses, e.g. contacts 1 – 5 (2 - 4) as voltage (current) leads. The 4W configuration uses contacts 2 - 3 (1- 4) as voltage (current) leads. Note that in the 4W(×) configuration, voltage electrodes are spaced apart by $200\mu m$ (several $10\mu m$). Furthermore, for the crossed configuration, the fit for both the CB and LL expressions almost coincide but only the LL fit gives a consistent $\gamma$

**value ($\gamma \leq 1$). Furthermore, the 4W (crossed) configuration yields $2\alpha_{bulk}$ ($\alpha_{end}$). Their ratio $\alpha_{end}/\alpha_{bulk} = 2.4 \pm 0.1$ can only be explained by LL theory.**

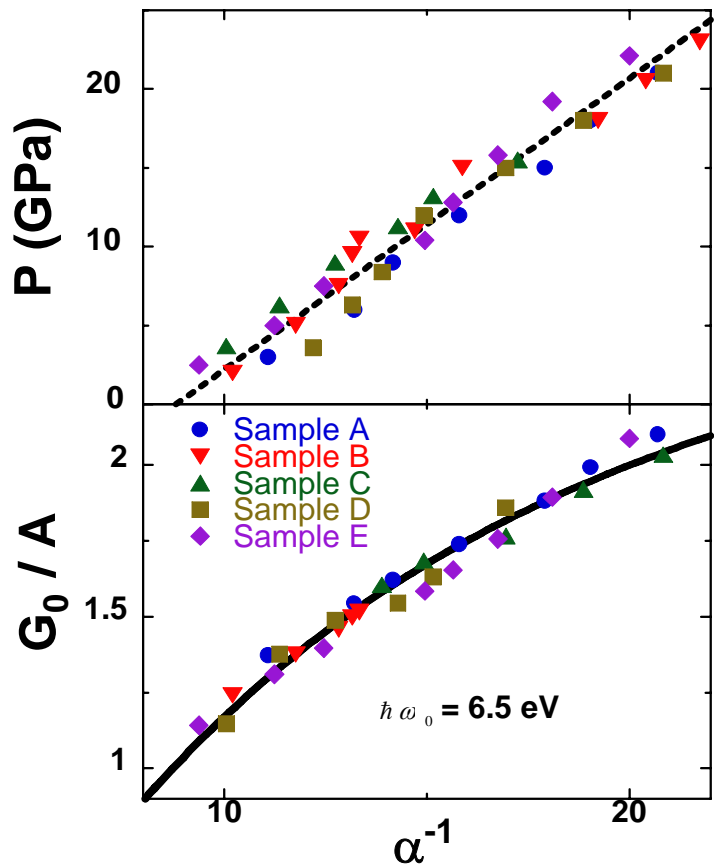

**Figure 3.** (Color online) Lower panel: scaling of the $G_0$ coefficient of the power law in temperature for the conductance as a function of the inverse $\alpha$ exponent. Upper panel: Pressure dependence of the inverse of the $\alpha$ exponent for all samples. We note that it is very similar for all the samples. The dashed line is the dependence considering a constant increase of doping under pressure (see text).